# A co-crystal between benzene and ethane, an evaporite material for Saturn's moon Titan.

Helen E. Maynard-Casely[a], Robert Hodyss[b], Morgan L. Cable[b] and Tuan Hoang Vu[b]

[a]Bragg Institute, Australian Nuclear Science and Technology Organisation, Locked Bag 2001, Kirrawee DC, 2232, Australia. [b]Jet Propulsion Laboratory, California Institute of Technology, 4800 Oak Grove Drive, Pasadena, California, 91109, USA.

## Abstract

Using synchrotron powder diffraction the structure of a co-crystal between benzene and ethane has been determined. The structure is remarkable, a lattice of benzene molecules playing host to ethane molecules. This is demonstrated by the similarity between the interactions found in the co-crystal structure and those in the pure structure, showing that the C-H…π network of benzene is maintained as a 'host' but expands to allow the ethane 'guest' to situate within the channels that result from this network. The co-crystal is determined to be a 3:1 benzene:ethane co-crystal and its structure is described by the trigonal space group $R\bar{3}$ with a = 15.977(1) Å and c = 5.581(1) Å at 90 K, resulting in a density of 1.067 g·cm$^{-3}$. Conditions under which this co-crystal forms indicate that it could readily be present on the surface of Saturn's moon Titan as an evaporite deposit following the evaporation of hydrocarbon fluids.

## Introduction

Crystallographic studies of benzene have a history of moving scientific understanding significantly forward. Kathleen Lonsdale's pioneering work on the structure of hexamethlybenzene showed the community that the benzene molecule was flat (Lonsdale, 1929), a study which, in part laid the ground work for molecular crystallography as we know it today. Later studies of the crystal structure of pure benzene (Cox *et al.*, 1958) showed that the flat benzene rings fit together like 'six-tooth bevel gear wheels' (Cox, 1932). The application of this fundamental knowledge has assisted in the solutions of thousands of structures that incorporate this parent aromatic compound.

There is now growing interest in the structures of many simple hydrocarbons for a very different application. Titan, Saturn's largest moon has in recent years been revealed, largely by the on-going Cassini mission, to have a 'hydrological' cycle. Unlike Earth's hydrological cycle, Titan's is not driven by water. On Titan the surface temperature is 91-95 K, and at these cryogenic temperatures the fluids that drive the cycle are small hydrocarbon molecules (Stofan *et al.*, 2007) such as methane and ethane, as well as dissolved nitrogen gas. Lakes and seas observed on the surface of Titan contain a mixture of methane and ethane (Cordier *et al.*, 2009), which result from cloud formation and precipitation in the atmosphere. Additionally there are a number of other small molecular species observed in the atmosphere, which are hypothesized to be present at the Titan surface. These include organic molecules such as hydrogen cyanide, acetylene, ethylene, acetonitrile, and benzene, formed photo-chemically from $CH_4$ and $N_2$ in the upper atmosphere (Vuitton *et al.*, 2008). In particular, benzene has been tentatively identified on the surface of Titan by the Huygens probe (Niemann *et al.*, 2005).

The observation of Titan's hydrological cycle now encompasses lakes, seas, clouds and even rain of small hydrocarbons – but it has been missing a vital piece. In light of the cycle observed at the Titan surface, it is natural to ask if Titan surface materials could produce deposits analogous to evaporites on Earth. Cassini imagery has collated evidence for possible evaporite deposits (Barnes *et al.*, 2011) , but it remains a mystery as to what these could be made of.  This is despite the important role such materials would play in both the hydrological cycle and the surface chemistry of Titan.

In light of the discovery of Titan's hydrological cycle, investigations have been undertaken to identify possible evaporite materials that would form on the surface of Titan.  Recent results with Raman and IR spectroscopy (Cable *et al.*, 2014)˙(Vu *et al.*, 2014) on the interaction of small molecules under Titan surface conditions identified the formation of a co-crystal between benzene and ethane. Although spectroscopy and quantum chemical calculations pointed to a specific local interaction between the ethane and benzene in the co-crystal, a crystallographic study was required to determine unambiguously the structure and consequently the composition of the co-crystal, as well ascertaining its viability as an evaporite material on Titan.

## Methods

Previous microscopic observations of the co-crystal showed that, on formation, the crystallite sizes were significantly reduced compared to that of frozen benzene (Vu *et al.*, 2014).  Hence, it was decided to pursue a powder-diffraction study of the potential co-crystal, using the high-resolution afforded by a synchrotron source.  Approximately 2 μL of benzene (Sigma Aldrich 99.8 %) was placed inside a 0.7 mm borosilicate capillary.  The amount of benzene was tailored so that the length of the drop within the capillary was ~2 mm. The capillary was then attached via a Swagelok fitting (Norby *et al.*, 1998) to a valve allowing the system to be closed.  This was then mounted on the powder diffraction beamline, Australian Synchrotron (Wallwork, 2007) , along with an Oxford Cryosystems cryostream (Cosier & Glazer, 1986) to control the temperature of the sample. The beamline was set up with $\lambda = 0.826(1)$ Å, verified from refinement of a pattern taken of a NIST LaB6 (SRM 660b) powder.  The X-ray beam from the synchrotron is vertically focused to a height of 1 mm, but as the source of the Powder Diffraction beamline is a bending magnet, there is no horizontal focussing.  Instead the width of the beam is constrained to 3 mm with lead slits, optimizing the instrument for data collection from capillary samples.

The benzene within the capillary was cooled to 130 K, and a diffraction pattern of this (and of all subsequent data collections) was obtained with a MYTHEN strip detector (Bergamaschi *et al.*, 2010) . Once the benzene was frozen, and the temperature of the cryostream and sample had reached 130 K, the position of the capillary was translated so that the edge of the frozen benzene aligned with the centre of the synchrotron X-ray beam.  The capillary system was then attached to a bottle of ethane (Sigma Aldrich 99.9 %).  With the temperature of the cryostream maintained at 130 K, ethane was condensed adjacent to the frozen benzene.  At these temperatures ethane is liquid, and very soon after opening the valve a drop of ethane condensed in contact with the frozen benzene.  This contact was monitored with X-ray diffraction, and after cycling the temperatures between 130 and 90 K diffraction peaks additional to those attributed to benzene were seen to form – interpreted to be from a benzene-ethane co-crystal.

This protocol results in rapid co-crystal formation, as verified by previous Raman experiments (see supplementary information). In these prior tests, liquid ethane was added to benzene frozen inside a 0.5 mm quartz capillary at 90 K. Upon warming to 130 K, co-crystal formation was observed within a few minutes, as evidenced by the characteristic Raman feature at 2873 cm$^{-1}$ [Vu *et al.*, 2014]. Once formed, these co-crystals remained stable inside the capillary and did not decompose until warmed above 135 K.

Once the co-crystal was formed at 130 K for the powder-diffraction study, the temperature of the cryostream was set to 90 K and the temperature stabilized there for a longer data acquisition. Acquiring a pattern at 90 K minimized thermal motion in the crystal structure and the longer acquisition revealed weaker peaks that may have been missed from the previous shorter acquisitions, these steps were undertaken to aid subsequent solution of the crystal structure of the new co-crystal.

Figure 1 contrasts the diffraction pattern of benzene taken at 170 K, with that of the contents of the capillary after ethane was condensed in contact with the frozen benzene, the temperature cycled and cooled to 90 K. In this pattern new diffraction peaks have appeared which cannot be attributed to benzene or solid ethane (which freezes at 89 K). To verify the result, the whole process of forming the co-crystal was then repeated and the diffraction from this also recorded.

To investigate the expansion of the co-crystal structure with heat, sequential patterns (collected over 120 seconds) were taken at 5 K intervals from 90 K to 150 K, then at 1 K intervals to 170 K. Previous results had suggested that the co-crystal would only be stable until ~160 K (Vu *et al.*, 2014), and the smaller temperature intervals allowed us to monitor the decomposition of the co-crystal.

## Results

### Determining the unit cell and symmetry

Working with the diffraction pattern taken at 90 K for 600 s, the diffraction peaks not attributed to benzene were fitted with pseudo-Voigt functions, with the TOPAS 4.1 program (supplied by Bruker-AXS) (Coelho, 2008). The positions of these peaks were then passed to the indexing routine within TOPAS to establish a unit cell to describe these positions. The highest ranked possibility of the output of this process was a trigonal cell with $a$ = 15.977(1) Å and $c$ = 5.581(1) Å, which gives a volume of 1233.77 Å$^3$. This unit cell was refined by Pawley refinement (Pawley, 1981), with the peaks described by the *R3* spacegroup (the lowest trigonal-symmetry spacegroup) and yielded a weighted R-factor (wRp) of 3.11 % and a goodness of fit (GoF) of 3.56.

Systematic absences of reflections of the trigonal indexing indicated that the symmetry of the co-crystal structure would conform to the $\bar{3}$ or $\bar{3}m$ Laue groups. Within these groups, there are five possible spacegroups: *R3*, *R$\bar{3}$*, *R32*, *R3m* and *R$\bar{3}m$*. In our analysis, the $\bar{3}m$ Laue class spacegroups (*R32*, *R3m* and *R$\bar{3}m$*) were ruled out because the additional mirror symmetry imposes too many restrictions on the positions of the benzene and ethane molecules.

After the indexing of the unit cell of the new co-crystal it was noted that there were still a number of small peaks that were not accounted for by this determination, shown in Figure 2. Further investigation of the data set indicated that these peaks persisted after the co-crystal melted, and seemed to be associated with diffraction from the residual benzene. Furthermore, these additional peaks were not observed during the second run of the experiment. Hence, these peaks were judged not to be from the co-crystal structure and were not considered further in the structure solution process. The first run was used for structure solution as the proportion of co-crystal formed (relative to the benzene) was higher.

### Determining the structure of the co-crystal

The volume of the unit cell of the co-crystal, 1233.77 Å$^3$ at 90 K, placed constraints on the likely contents and stoichiometry of the structure. The density of ethane in its solid phase at 89 K is 0.669 gcm$^{-3}$ (Van Nes & Vos, 1978) and benzene at 90 K is 1.103 g·cm$^{-3}$ (Bacon *et al.*, 1964), so it would seem reasonable to assume that the density of the co-crystal would lie between these. However, the indexing of the cell to a trigonal group also placed further constraints on the contents. For instance, in order for a 2:1 benzene and ethane co-crystal to conform to the three-fold symmetry there would have to be three or six formula units within the unit cell. This would lead to densities for the co-crystal of 0.752 (for three formula units) or 1.504 g·cm$^{-3}$ (for six formula units).

Table 1 – Possible density values of the unit cell contents for the benzene:ethane co-crystal. Values given are the calculated density (in g·cm$^{-3}$) of the material with the respective benzene:ethane ratio and number of formula units within the cell. The coloured cells indicate those values that conform to the density (yellow) and symmetry (green) constraints have been applied at this stage.

|     | 2 | 3 | 4 | 5 | 6 | 7 |
| --- | --- | --- | --- | --- | --- | --- |
| 1:1 | 0.291 | 0.437 | 0.582 | 0.728 | 0.873 | 1.019 |
| 1:2 | 0.372 | 0.558 | 0.744 | 0.930 | 1.116 | 1.302 |
| 1:3 | 0.453 | 0.680 | 0.906 | 1.133 | 1.359 | 1.586 |
| 2:1 | 0.501 | 0.752 | 1.003 | 1.253 | 1.504 | 1.755 |
| 3:1 | 0.712 | 1.067 | 1.423 | 1.779 | 2.135 | 2.491 |

allowed from density constraints
allowed from symmetry constraints

In fact, as Table 1 shows, using the justification that the density of the co-crystal would be between that of benzene and ethane, there are only three possible contents identified by these means. These are a 1:1 benzene-ethane co-crystal with six formula units, a 2:1 co-crystal with three formula units, and a 3:1 co-crystal also with three formula units. Unusual circumstances (i.e. the density of the co-crystal being lower than that of ethane or higher than that of benzene) would have been pursued, if the structure solution using these potential contents had not been successful.

Using the identified potential unit-cell contents, attempts were made to establish the atomic positions of the benzene and ethane molecules within the co-crystal. To aid this structure solution process (and to minimize the variables), rigid-body units describing the benzene and ethane molecules were constructed, details of which are given in the Supplementary information. Once the rigid-body units had been constructed, these were entered into the Free Objects for Crystallography (FOX) program version 9.1 (Favre-Nicolin & Cerny, 2002) within the constraints of the trigonal unit cell determined during the indexing process. The arrangement of the rigid-body units were minimized against the 90 K pattern, presented in Figure 1, with parallel tempering. Prior to the minimization, a Le Bail refinement (Le Bail, 2005) of benzene was also undertaken in FOX, to account for the peaks from this material in the pattern, the results of which were added to the parallel tempering calculation. This process was undertaken for each of the three possible contents identified and in both of the two spacegroups (*R3* and *R$\bar{3}$*) identified in the indexing procedures.

From this process a viable crystal structure (judged by bond distances and fit to the observed data) was obtained only for the 3:1 benzene:ethane co-crystal, with three formulae units within the cell. This structural model was prepared for Rietveld refinement in the TOPAS 4.1 program.

**Refinement of the 3:1 benzene:ethane co-crystal**

Figure 3 shows a Rietveld fit (Rietveld, 1969) of the presented structure to the 90 K data. The parameters varied for this refinement were a scale factor, a background function, a zero error to account for displacement of the capillary from the beam centre, a single broad peak to account for the scattering of the borosilicate capillary, the lattice parameters for both of the phases, and the orientation and translation of the two rigid units used to build the co-crystal structure (details of these are given in the Supplementary information). Peak-shape parameters (Thompson-Cox-Hasting model (Thompson *et al.*, 1987) ) determined from the Pawley refinement were used, but fixed for the structural refinement, with crystallite size refined. Additional pseudo-Voigt peaks were also entered into the refinement to account for the intensity of the small peaks that were revealed at the indexing stage, as shown in Figure 2. In order not to over-parameterize the refinement of this structure, atomic displacement parameters were constrained to be the same for each atom type (i.e. one for carbon atoms in the benzene, one for the carbon atoms in the ethane, etc.).

Table 2 lists the atomic fractional coordinates that describe the structure solution of a 3:1 co-crystal of benzene and ethane. The resultant density of this phase is 1.067 g·cm$^{-3}$ at 90 K, slightly less dense than solid benzene at 90 K of 1.103 g·cm$^{-3}$. The hydrogen-atom positions that we have determined come solely from geometric placing within the rigid bodies that were generated to solve the structure.

Table 2 - Atomic fractional coordinates of the co-crystal model from the refinement presented in Figure 3.

| Atom | x | y | z | $U_{iso}$ (Å$^2$) |
|---|---|---|---|---|
| C1b | 0.192(2) | 0.261(2) | 0.376(2) | 2.85(3) |
| C2b | 0.212(2) | 0.332(2) | 0.559(3) | 2.85(3) |
| C3b | 0.189(2) | 0.405(1) | 0.511(2) | 2.85(3) |
| H1b | 0.213(2) | 0.206(2) | 0.403(2) | 0.5(4) |
| H2b | 0.248(2) | 0.332(2) | 0.724(3) | 0.5(4) |
| H3b | 0.203(2) | 0.459(1) | 0.648(2) | 0.5(4) |
| C1e | 0.0000 | 0.0000 | 0.376(2) | 2.49(11) |

The model of the benzene:ethane co-crystal that has been determined is a relatively static one. From these data potential disorder within the structure cannot be ruled out; however the fit of the model to the data would suggest that this would be limited in its extent. Additional conformation of the structure was sought using Fourier difference methods to determine if there was any systematic electron density not accounted for by the structure. Structure factors were extracted from the refinement presented in Figure 3, and a $F_{(obs)}$-$F_{(calc)}$ calculation performed in the VESTA program (Momma & Izumi, 2008). The only significant feature in this difference map was located along the *c*-axis, and could explain the larger displacement parameters determined from refinement for the hydrogen atoms attached to the ethane molecule.

The presented structure was confirmed with the same refinement procedure undertaken on the pattern collected from the second formation of the co-crystal at 90 K. Details of this fit are presented in the supplementary information.

**Description of the benzene:ethane co-crystal**

At first glance the structure of the co-crystal that has been determined can be described as a two-dimensional host-guest material. The benzene molecules are arranged in a ring of six molecules around each of the ethane molecules. This creates channels through the structure where the ethane molecules reside. Inspection of the co-crystal over a number of unit cells shows that the structure is maintained by a network of C-H…π interactions very similar to those found in crystalline benzene.

Figure 4 compares the benzene crystal structure (as determined by Bacon *et al. (Bacon et al., 1964)*) with the benzene:ethane co-crystal. This shows that the ethane molecules being surrounded by six benzene molecules is a feature 'inherited' from the pure benzene structure. In the formation of the co-crystal the ethane molecules have replaced a quarter of the benzene molecules. This replacement by a smaller molecule has resulted in the benzene molecules in the co-crystal rotating into higher symmetry (6-fold) positions. Crucially the benzene molecule maintain very similar C-H…π interactions (as shown by the distances highlighted in Figure 4) compared to those in pure benzene.

The similarities between the benzene:ethane co-crystal and pure benzene are explored further in Figure 5. Additionally, the structure is compared with the only other co-crystal between benzene and a small hydrocarbon, a 1:1 benzene:acetylene co-crystal (Boese *et al.*, 2003). This shows how C-H…π interactions between the benzene molecules create planes that run through both the benzene:ethane co-crystal and benzene structure. The higher symmetry of the benzene:ethane co-crystal means that the benzene interactions create the channels parallel to the *c* axis where the ethane molecules are situated. The interactions in the benzene:ethane co-crystal and pure benzene are in stark contrast to the inter-molecular interaction in the 1:1 benzene:acetylene co-crystal. Here the benzene molecules are arranged such that they do not interact each other, instead are seen to form C-H…π bonds with the acetylene molecules which are aligned perpendicular to the benzene rings.

Though the planes of the interactions are parallel to the *c* axis, the chains of C-H…π interactions are inclined to this, and in differing directions to each other. This has implications for the thermal properties of the co-crystal as explained in the proceeding section. The structure indicates that there is only limited interaction between the benzene chains and the ethane molecules. This is in contrast to prior interpretations of the co-crystal structure based on Raman spectroscopy and quantum mechanical modelling (Vu *et al.*, 2014), which suggested a C-H…π interaction between ethane and benzene. This discrepancy is investigated in the discussion section.

The structure of the benzene:ethane co-crystal indicates that the benzene molecules form a two-dimensional host material, with the ethane molecules as guests. Like the three-dimensional gas clathrate system, this raises the potential for further guest species of similar dimensions to the ethane molecule also to form co-crystals with benzene.

**Expansion and stability of the co-crystal with temperature**

Using the model of the co-crystal determined by the structure solution process, each of the patterns from 90 – 165 K was refined and the lattice parameters and proportion of each phase (either benzene or co-crystal) extracted from each pattern.

Figure 6 a) presents how the lattice parameters vary over the temperature range. From this, it can be seen that the co-crystal structure exhibits significant anisotropic thermal expansion, with the majority of the expansion occurring in the *a* and *b* axes. This is interpreted as resulting from a consequence of the C-H… π interactions which are aligned closer to the *a* and *b* axes in the co-crystal structure. In the *c* axis these chains of interactions interlock with each other and restrict the expansion in this direction.

The stability of the co-crystal structure is demonstrated by Figure 6 b) which charts the relative proportion of benzene and co-crystal refined in each diffraction pattern collected, the patterns taken between 150 and 170 K are presented in Figure 6 c). The proportion of co-crystal to benzene in the patterns is steady from 90 K at ~89 % co-crystal and ~11 % benzene until 145 K. Above this temperature, the proportion of the co-crystal in the refined pattern decreases monotonically. This is consistent with previous observations of the decomposition of the co-crystal at elevated temperatures (Cable *et al.*, 2014), (Vu *et al.*, 2014). It is likely that, given sufficient time, above 145 K the proportion of the co-crystal structure would decompose without any increase of temperature.

## Discussion

The intermolecular interactions between the ethane and benzene molecules in the co-crystal structure are quite different to those considered by Vu *et. al.* (Vu *et al.*, 2014). In this previous work the electrostatic potential surfaces of three benzene-ethane dimers were used to rationalize the origin of the Raman shifts observed for the benzene and ethane features in the spectra from the co-crystal. The three dimers were characterized by having one, two, or three C-H bonds pointing towards the centre of the benzene ring – mono-, bi-, or tridentate. The most significant mode was the $v_{11}(e_g)$ $CH_3$ deformation stretch vibrational mode in the ethane. In the co-crystal this modes has a frequency at 90 K of 1454.8 cm$^{-1}$, giving a Raman shift ($\Delta \tilde{v}$) of -12.3 cm$^{-1}$ compared to the liquid at 1467.1 cm$^{-1}$. The previous work found the largest shift for this mode among the dimers studied to be a $\Delta \tilde{v}$ of -7 cm$^{-1}$ arising from a monodentate interaction between the benzene and ethane molecules. The interaction between the benzene and ethane molecules is decidedly weaker than the monodentate interaction, and the geometry is unfavourable for a C-H…π interaction between ethane and benzene (as can be seen in Figure 4). The red shift of the Raman peak relative to the liquid suggests that the ethane molecule is in a less confined geometry compared to in the liquid – and this is consistent with the structure where the ethane molecules are situated in channels formed by the lattice of benzene molecules.

The observed change in the Raman shifts of the ethane molecules in the co-crystal from their pure form at 90 K (of -12.2 cm$^{-1}$ for the $v_1(a_{1g})$ and -12.3 cm$^{-1}$ for the $v_{11}(e_g)$ mode) contrasts starkly with those shifts observed in the benzene molecules, 0.3 cm$^{-1}$ for the $v_1$; -3.1 and -1.8 cm$^{-1}$ for the $v_7$. The structure of the co-crystal explains this, as the intermolecular interactions between the benzene molecules are largely unchanged compared to those found in pure solid benzene, as illustrated by Figures 4 and 5. All of the above discussion leads to the conclusion that the guest ethane molecules within the co-crystal are only weakly bonded (most likely through van der Waals interactions) with host benzene molecules.

Figure 7 demonstrates the inherent weakness of the C-H…π interaction as the dominant intermolecular interaction, compared to that of other possible Titan 'minerals' where the hydrogen bond dominates these interactions (Fortes *et al.*, 2003, Belosludov *et al.*, 2002). Both pure benzene and the co-crystal structure show significantly higher relative thermal expansion compared to that of methane clathrate and ammonia dihydrate. This suggests that the crystal structures of benzene and the co-crystal are held together by weaker intermolecular forces, and more likely to be affected significantly by temperature.

How does the knowledge of the structure change our understanding of the co-crystal's formation? Observations of this showed that the morphology of the benzene changes substantially (Vu *et al.*, 2014), specifically that the crystals were seen to break into smaller crystallites. This is consistent with the determination of the structure, with a quarter of the benzene molecules being replaced by ethane. This causes a significant change in the arrangement of the benzene molecules, forcing them to move to take up higher symmetry positions. It would seem this change is enough to put additional strain through the crystal structure, which could be released by the breaking into smaller crystallites.

It seems likely, given the channels that the ethane molecules are found in within the benzene:ethane co-crystal, that other guest species could form similar co-crystals with benzene. Exchange and partial substitution of ethane with other linear hydrocarbons, such as hydrogen cycanide (Cordier *et al.*, 2013), will also be investigated further enriching the possibilities of Titan's icy mineralogy.

## Conclusions

This work has confirmed that a co-crystal is able to form between benzene and ethane, not only by showing the unique diffraction pattern of this new material but by also solving its atomic arrangement using synchrotron powder diffraction. Furthermore, it is shown that this benzene:ethane co-crystal is a viable candidate to be an evaporite material on the surface of Saturn's moon Titan, as exhibited by the fact the structure is maintained over a range of temperatures. It is also the first potential 'icy mineral' to be identified where its intermolecular interactions are not dominated by hydrogen bonding. The weaker C-H…π interactions which stabilise the benzene:ethane co-crystal means that this can be presented as the first of new group of materials, with similar physical properties, that could shape the surface of Titan.

The structure of the benzene:ethane co-crystal is substantially different from any known co-crystal of benzene. It is in significant contrast to co-crystals formed between benzene and acetylene (Boese *et al.*, 2003), where the linear acetylene molecules align normal to the benzene rings. The similarity of the interactions between the co-crystal and the benzene structure show that the C-H…π network of benzene is maintained as a 'host' but expands to allow the ethane 'guest' to situate within the channels that result from this network.

We anticipate that this work will be followed by a number of other investigations charting the co-crystal formation and stability of other small molecular species that would become Titan 'minerals' and contribute to the understanding of the geology and potential habitability of this icy moon.

## Acknowledgements

The authors wish to acknowledge the Australian Synchrotron for the award of beamtime EPN 3200 and Justin Kimpton for his assistance during this experiment. RH acknowledges the support of the NASA Astrobiology Institute (Titan as a Prebiotic Chemical System) and NASA's Outer Planets Research program. Part of this research was carried out at the Jet Propulsion Laboratory, California Institute of Technology, under a contract with the National Aeronautics and Space Administration.

Additionally supplementary material is provided detailing the formation of the co-crystal as observed by Raman spectroscopy, refinement of a second formation of the co-crystal and the structure and refined data are presented in Crystallographic Information File (CIF) format.

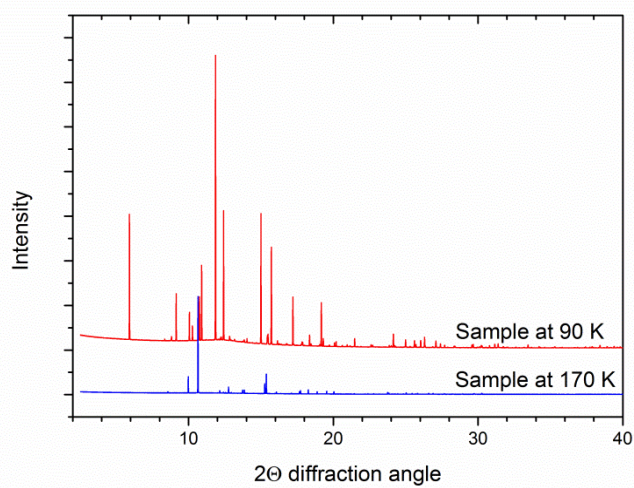

Figure 1 – Comparison of the diffraction pattern of the sample at 90K and that at 170 K. The pattern at 90 K contains diffraction from both the co-crystal and benzene, and the pattern at 170 K only contains diffraction from benzene.

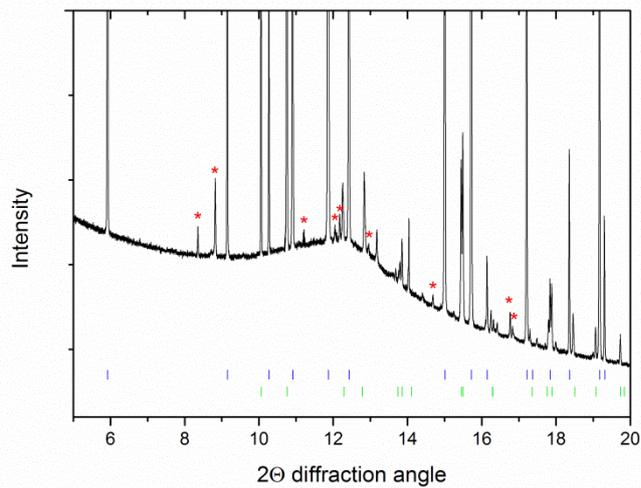

Figure 2 – Details of the pattern of the co-crystal and benzene, presented in Figure 1. The blue tick marks indicate the expected positions of Bragg peaks from the $R\bar{3}$ unit cell of the co-crystal, and the green tick marks indicate the expected positions of the benzene diffraction. As explained in the text, there were a number of residual peaks, indicated by red asterisks.

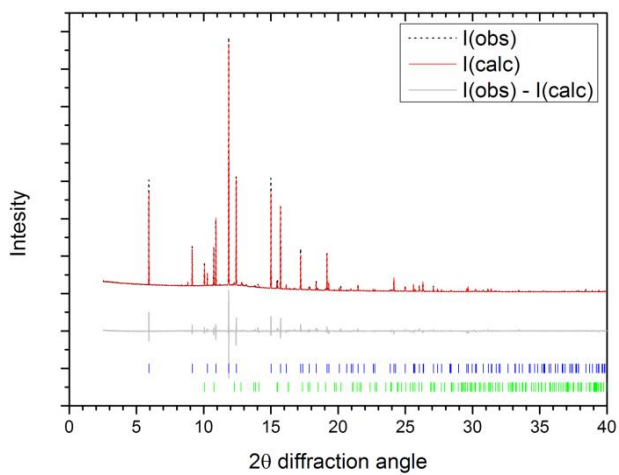

Figure 3 - Rietveld fit of the 90 K sample pattern with the structural parameters of benzene and the co-crystal structure determined in this work. This gives a wRp of 5.33 % and a GoF of 4.77. The grey line below the data indicates the difference between the observed and calculated patterns. The blue tick marks indicate the positions of reflections from the co-crystal structure and green tick marks indicate the positions of reflections from benzene (Cox, 1932).

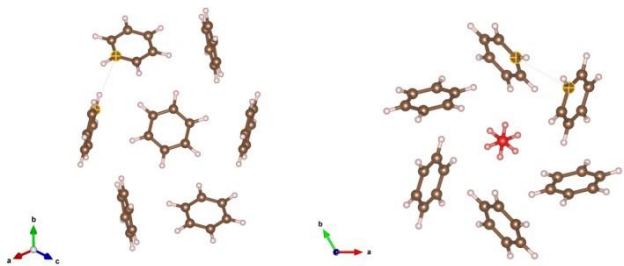

Figure 4 A section of the benzene structure(Bacon *et al.*, 1964) (left) compared with the structure of the benzene:ethane co-crystal determined by this work(right). The benzene structure is viewed down the [111] direction and the highlighted carbon – carbon distance is 3.855 Å. The representation of the benzene:ethane co-crystal structure is viewed down the *c* axis and the ethane molecule is coloured red to distingish it. The highlighted carbon-carbon distance in the co-crystal representation is 3.836 Å.

Benzene:ethane co-crystal

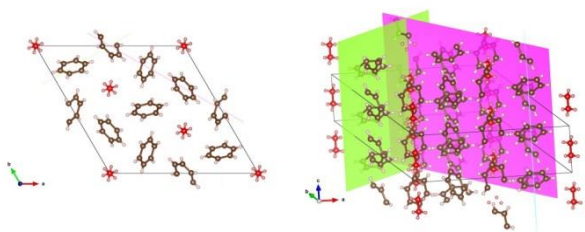

*Pbca* Benzene

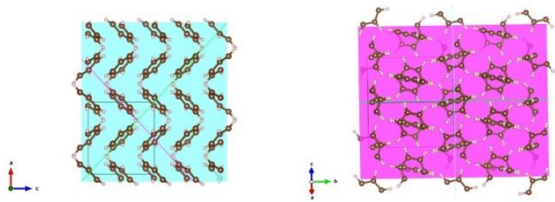

Benzene:acetylene co-crystal

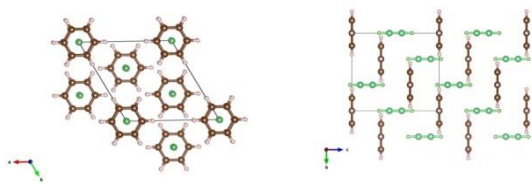

Figure 5 The packing of molecules and planes of C-H…π interactions across the benzene:ethane co-crystal determined in this work, compared with that of benzene (Bacon *et al.*, 1964) and the benzene acetylene co-crystal (Boese *et al.*, 2003). In the representation of the benzene:ethane co-crystal the ethane molecules are coloured red to distinguish them from the benzene molecules, similarly in the 1:1 benzene:acetylene co-crystal the acetylene molecule are coloured green for the same purpose. The benzene:ethane and benzene structure representations have the planes of hydrogen - π interactions highlighted.

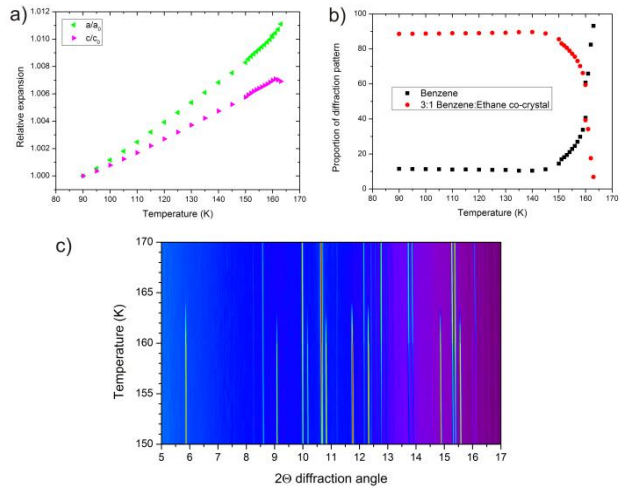

Figure 6 a) The relative expansion of the a and c axes over the temperature range studied of the co-crystal normalised to 90 K.  b) The relative proportion of the co-crystal and solid benzene in the patterns refined. c) shows a plot of the diffraction data over the temperature range that the co-crystal decomposes.

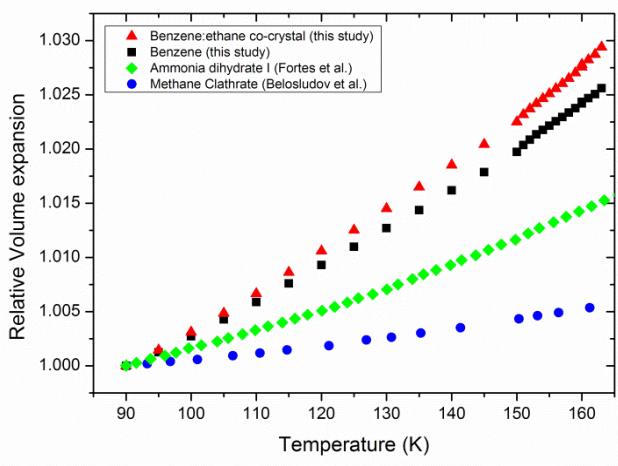

Figure 7 – comparison of the relative thermal expansion (normalized to 90 K) of the benzene:ethane co-crystal with other materials, pure benzene (where the intermolecular interactions are dominated by C-H…π interactions) and ammonia dihydrate I (Fortes *et al.*, 2003) and methane hydrate I (Belosludov *et al.*, 2002) (where the interactions within the crystal structure are dominated by hydrogen bonding).